\documentclass[twoside,twocolumn,9pt]{article}
\usepackage{extsizes}
\usepackage[super,sort&compress,comma]{natbib} 
\usepackage[left=1.75cm, right=1.75cm, top=2cm, bottom=2.0cm]{geometry}
\usepackage{balance}

\usepackage{sectsty}
\allsectionsfont{\scshape}

\usepackage{graphicx} 
\usepackage{lastpage}
\usepackage{float}
\usepackage{fancyhdr}
\usepackage{fnpos}
\usepackage[english]{babel}
\usepackage{array}
\usepackage[T1]{fontenc}
\usepackage[usenames,dvipsnames]{xcolor}
\usepackage{setspace}
\usepackage{hyperref}
\usepackage[squaren]{SIunits}

\usepackage{upgreek}
\usepackage{xspace}
\usepackage{placeins}
\usepackage{amsmath,amssymb}

\begin{document}

\twocolumn[
  \begin{@twocolumnfalse}
\vspace{3cm}

\begin{center}

    \noindent\huge{\textbf{\textsc{Particulate suspension coating of capillary tubes}}} \\
    \vspace{1cm}

    \noindent\large{D.-H. Jeong,\textit{$^{\,a}$} L. Xing,\textit{$^{\,a}$} J.-B. Boutin,\textit{$^{\,a}$} and Alban Sauret \textit{$^{a}$}}$^{\ast}$ \\

    \vspace{5mm}
    \noindent\large{\today} \\

    \vspace{1cm}
    \textbf{\textsc{Abstract}}
    \vspace{2mm}

\end{center}

\noindent\normalsize{The displacement of a suspension of particles by an immiscible fluid in a capillary tube or in a porous media is a canonical configuration that finds application in a large number of natural and industrial applications, including water purification, dispersion of colloids and microplastics, coating and functionalization of tubings. The influence of particles dispersed in the fluid on the interfacial dynamics and on the properties of the liquid film left behind remain poorly understood. Here, we study the deposition of a coating film on the walls of a capillary tube induced by the translation of a suspension plug pushed by air. We identify the different deposition regimes as a function of the translation speed of the plug, the particle size, and the volume fraction of the suspension. The thickness of the coating film is characterized, and we show that similarly to dip coating, three coating regimes, liquid only, heterogeneous, and thick films, are observed. We also show that, at first order, the thickness of films thicker than the particle diameter can be predicted using the effective viscosity of the suspension. Nevertheless, we also report that for large particles and concentrated suspensions, a shear-induced migration mechanism leads to local variations in volume fraction and modifies the deposited film thickness and composition.} \\

 \end{@twocolumnfalse} \vspace{0.6cm}

  ]

\makeatletter
\renewcommand*{\@makefnmark}{}
\footnotetext{\textit{$^{a}$~Department of Mechanical Engineering, University of California, Santa Barbara, California 93106, USA}}
\footnotetext{\textit{$^{*}$ asauret@ucsb.edu}}
\makeatother

\section{Introduction} \label{sec:intro}

The displacement of a suspension of particles by another immiscible fluid (air or liquid) in confined geometries and porous media is an important process involved in many industrial and natural situations. For instance, such processes are encountered in enhanced oil recovery,\cite{orr1984use} infiltration of liquid in porous media,\cite{cueto2008nonlocal} the transport and pollution by microplastic in soils,\cite{alimi2018microplastics,wu2021film} the intermittent dispensing of liquids, and microfluidics\cite{whitesides2006origins,dressaire2017clogging} among other situations. This configuration can also be leveraged to develop new coating processes to functionalize tubings.\cite{primkulov2020spin} Besides, a common approach to model multiphase flow in porous media is to consider the simplified system made of capillary tubes.\cite{olbricht1996pore} Therefore, a fundamental understanding of the dynamics observed in this configuration for a large variety of fluids encountered in practical applications is required.

When a wetting liquid plug is pushed in a capillary tube by an immiscible fluid, \textit{e.g.}, air, it deposits a thin film of thickness $h$ on the wall of the tube.\cite{taylor1961deposition,reinelt1985penetration,giavedoni1997axisymmetric,aussillous2000quick,jalaal2016long} For a homogeneous Newtonian liquid of dynamic viscosity $\eta_{\rm f}$ and interfacial tension $\gamma$ pushed by air, the formation of the coating film is governed by the competition between viscous and surface tension forces through the capillary number ${\rm Ca}=\eta_{\rm f}\,U/\gamma$, where $U$ denotes the velocity of the moving air/liquid interface at the rear of the plug. The thickness $h$ of the liquid film deposited on the wall of a tube of radius $R$ is a function of $R$ and ${\rm Ca}$ only.\cite{aussillous2000quick,balestra2018viscous} If the fluid is partially wetting, \textit{i.e.}, its contact angle is $\theta > 0$, the film dynamics is more complex as the contact line motion plays a role in determining the deposition patterns.\cite{zhao2018forced,hayoun2022triple}

\begin{figure*}
\centering
\includegraphics[width = \textwidth]{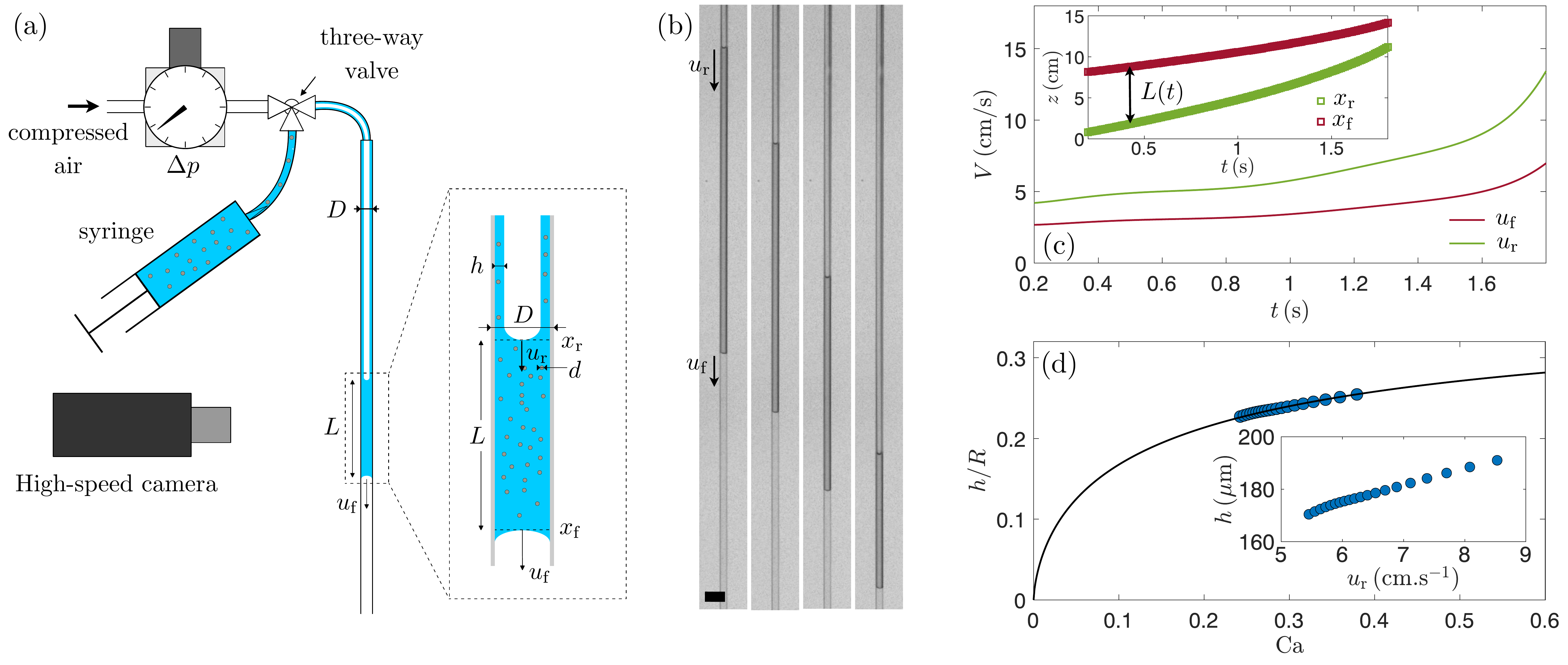}
\caption{(a) Schematic of the experimental setup used to study the displacement of a suspension plug and the deposition of a coating film in a circular capillary tube of diameter $D=2\,R$. The capillary tube is connected to the atmosphere on one end, and the suspension plug is pushed by applying a constant pressure difference $\Delta p$. (b) Example of an experiment showing that the translating plug leaves behind a coating film of thickness $h$. The time goes from left to right, and the scale bar is $2\,{\rm  mm}$. (c) Example of measurement of the velocity of the front $u_{\rm f}$ and of the rear $u_{\rm r}$  of the plug from the time evolution of their respective position $x_{\rm f}$ and $x_{\rm r}$ (shown in inset). (d) Resulting coating thickness rescaled by the radius of the capillary $h/R$ calculated using the measurements presented in (c) and Eq. (\ref{eq:method_Aussillous}). The solid line is given by Eq. (\ref{eq:Taylor}). Inset: Coating thickness $h$ as a function of the velocity of the rear of the plug $u_{\rm r}$.}
\label{fig:setup}
\end{figure*}

When solid particles are dispersed in a liquid, the usual interfacial dynamics and rheological approaches used to describe capillary flows can often fail when the lengthscale of the capillary object, here the liquid film, becomes comparable to the diameter of the particles. This peculiar dynamics of suspension has been considered in various configurations such as the formation of droplets,\cite{furbank2004experimental,bonnoit2012accelerated,chateau2018pinch,thievenaz2021pinch,thievenaz2021droplet,thievenaz2022onset} the stability of jets,\cite{chateau2019breakup} the fragmentation of suspension sheets,\cite{raux2020spreading} the motion of contact lines,\cite{zhao2020spreading} and during the formation of thin films.\cite{gans2019dip} The challenge arises from the fact that for capillary flows of suspensions, in addition to the film thickness $h$, an additional lengthscale enters into the problem: the particle diameter $d$. This complexity when using a suspension of non-Brownian particles dispersed in non-volatile liquids has been considered during the formation of a thin film of suspension in a dip-coating process.\cite{gans2019dip,palma2019dip} The dip-coating of a substrate consists of withdrawing a solid body initially dipped in a liquid bath so that it emerges covered with a thin layer of liquid.\cite{ruschak1985coating,rio2017withdrawing} The thickness $h$ of liquid deposited on a plate withdrawn at a velocity $U$ from a bath of Newtonian fluid of viscosity $\eta_{\rm f}$ and surface tension $\gamma$ follows the Landau-Levich-Deryaguin law (LLD): $ h = 0.94 \, \ell_c \, {\rm Ca}^{2/3} $ (for ${\rm Ca}<10^{-2}$)  where ${\rm Ca}= \eta_{\rm f} \, U / \gamma $ is the capillary number and $ \ell_c = \sqrt{\gamma/(\rho\,g)}$ is the capillary length.\cite{landau1942physicochim,Deryagin,Quere:1999wv,Maleki:2011in,rio2017withdrawing} The dip coating with suspensions of non-Brownian particles has been shown to exhibit different coating regimes depending on the particle diameter $d$ and the coating thickness.\cite{gans2019dip,palma2019dip,jeong2022dip} At small capillary number ${\rm Ca}$, corresponding to very thin films ($h \ll d$), no particles are entrained, and only a liquid film coats the plate. At intermediate capillary numbers leading to $h \lesssim d$, a heterogeneous coating made of clusters of particles arranged as a monolayer is observed on the substrate. Finally, at large capillary numbers where $h > d$, the thickness of the entrained film is captured by the Landau-Levich-Derjaguin law using the effective viscosity of the suspension $\eta(\phi)$, where $\phi$ is the solid fraction of particles in the suspension. A similar behavior has also been reported for the dip coating of cylindrical substrates when accounting for the change in the LLD law due to the curvature of the substrate.\cite{dincau2020entrainment} The role of the particle diameter, compared to the film thickness, has allowed the development of new capillary filtration\cite{sauret2019capillary} and sorting methods\cite{dincau2019capillary,jeong2022dip} leveraging the dip coating platform.

Interestingly, the fluid mechanics and flow topology underlying the dip coating process share many common features with the flow of a plug of liquid in a tube, including the presence of a stagnation point governing the coating film and the evolution of the film thickness $h$ with the capillary number ${\rm Ca}$.\cite{krechetnikov2010application} As a result, the entrainment threshold of isolated particles in the coating film during the dip coating of a plate is consistent with the entrainment threshold in a capillary tube.\cite{sauret2019capillary,jeong2020deposition} However, the relevance of the different coating regimes and the properties of the coating film remain more elusive for a plug of particulate suspension at a moderate volume fraction pushed by an immiscible fluid in cylindrical capillary tubes. Yet, such an approach could allow coating the inner wall of tubes, giving them some surface properties.\cite{hayoun2018method,primkulov2020spin} It would also lead to a better description of multiphase flows that involve particles and interfaces in porous media.

In this article, we focus on the intermittent flow of a plug of non-Brownian and neutrally buoyant suspension of monodisperse spherical particles in cylindrical capillaries. We experimentally characterize the deposition of the coating film during the process and illustrate how the nature of the coating depends on the velocity $U$ of the plug of suspension for different particle diameters $d$. We first present our experimental methods in section \ref{sec:Experimental}. Section \ref{sec:results} reports our general observations. We consider in section \ref{sec:discussion} the nature of the coating film in more detail, and we investigate the role of the particle diameter and of the volume fraction of the suspension. We also discuss the important role of shear-induced migration to rationalize our observations.


\section{Experimental methods} \label{sec:Experimental}

\subsection{Experimental setup}

The experimental setup, presented in figure \ref{fig:setup}(a), consists of a cylindrical glass capillary tube connected to compressed air, and to a syringe. The glass capillary tube (Vitrocom) is $30\,{\rm cm}$ long and has an inner diameter $D=2\,R=1.5\,{\rm mm}$. The results with different diameters can be rescaled with the diameter of the capillary tube so that we focus here on one diameter and vary the other parameters. The only constraint is that the diameter of the particles should be small compared to the diameter of the tube.\cite{jeong2020deposition} Similarly, the viscosity of the interstitial fluid can be rescaled through the capillary number and will not modify the dynamics,\cite{gans2019dip} therefore we kept the same interstitial fluid throughout the study.

The suspensions are made of spherical polystyrene particles (Dynoseeds TS, Microbeads) of diameter $20\,{\rm \mu m} \leq d  \leq 250 \,{\rm \mu m}$ dispersed in a high-density silicon oil (Sigma Aldrich). The silicone oil has a dynamic viscosity at $20^{\rm o}{\rm C}$ of $\eta_{\rm f}=0.11\,{\rm Pa\,s}$, an interfacial tension $\gamma \simeq 25 \pm 2\,{\rm mN\,m^{-1}}$ and a density $\rho_{\rm p} = 1058\, {\rm kg\,m^{-3}}$. The silicone oil perfectly wets both the capillary tube and the particles.\cite{bonnoit2012accelerated} The densities of the particles range between $\rho_{\rm p} = 1056\, {\rm kg\,m^{-3}}$ and $1062\, {\rm kg\,m^{-3}}$ depending on the batch considered. A mechanical stirrer is used when initially preparing the suspension and during a series of experiments to disperse the particles in the silicone oil. The average density of the polystyrene particles is close enough to the density of the interstitial fluid, and the capillary tube is placed vertically, so that the suspensions can be considered neutrally buoyant over the timescale of one experiment, typically from a few seconds to a few tens of seconds. The volume fraction of the suspension, $\phi=V_{\rm p}/V_{\rm tot}$, where $V_{\rm p}$ and $V_{\rm tot}$ are the volume of particles and the total volume of suspension, respectively, is varied in the range $0.05<\phi<0.25$. In this study, we did not consider larger volume fractions as the pressure required to translate the plug at a significant speed would be too large for our system. In addition, for volume fractions larger than $\phi=0.35$ we observed the apparition of an unstable finger at the front of a suspension plug, which could modify the dynamics.\cite{kim2017formation}

To visualize the motion of the plug, we use a $20\times20$ cm backlight LED Panel (Phlox), and the dynamics is recorded with a high-speed camera (Phantom VEO 710L) equipped with a macro-lens (Nikkor 200mm). Additional photos of the coating films are taken with a Nikon D7200 camera and a 200 mm macro lens. After each run, the capillary tube is thoroughly cleaned with Isopropanol (IPA, Sigma Aldrich), rinsed with DI water (Millipore), and dried with compressed air before being reused.


\subsection{Measurement of the coating thickness}

At the beginning of an experiment, a suspension plug of volume $V$ is placed at the inlet of the tube using the syringe. We ensure that the initial length of the plug is about $L \sim 10\,{\rm cm}$, so that $L>>2\,R$. The inlet of the tube is then closed by a three-way valve, and the plug is initially at rest. The entrance of the tube is connected to a source of compressed air with the three-way valve, as shown in figure \ref{fig:setup}(a). The plug is set in motion by opening the entrance of the setup to the pressurized chamber. The pressure is controlled using a pressure regulator (Omega AR91-015) in the range $350 \,{\rm Pa}<\Delta p < 30,000 \,{\rm Pa}$. The suspension plug then translates in the capillary tube at a typical velocity between a few ${\rm mm/s}$ to a few ${\rm cm/s}$, as shown in the example of figure \ref{fig:setup}(b). The Reynolds number based on the radius of the capillary remains smaller than 0.1, and we can neglect inertial effects. Another relevant dimensionless parameter in this system is the Weber number, which compares inertia with the capillary force and is defined as ${\rm We}={\rho {u_{\rm r}}^2\,(R-h)}/{\gamma}$,\cite{aussillous2000quick}. The Weber number remains smaller than $0.2$ in our experiments, meaning that inertia effects are also negligible compared to capillary effects. The length of the plug decreases over time as the liquid is continuously deposited at the rear of the plug on the capillary wall. As a result, since the translation of the plug is driven by a pressure difference, the translational velocity increases during one experiment. We measure the instantaneous velocity and the corresponding local thickness, following the method developed by Aussillous \& Qu\'er\'e for Newtonian liquids.\cite{aussillous2000quick}

To measure the thickness of the coating film, we rely on the measurement of the time-evolution of the position of the front and rear interfaces, $x_{\rm f}$ and $x_{\rm r}$, respectively.\cite{aussillous2000quick} The positions of the menisci are extracted from the movies using ImageJ and custom-made Matlab routines. An example is shown in the inset of figure \ref{fig:setup}(c). The length of the plug, $L(t)=\mid x_{\rm f}-x_{\rm r} \mid$ decreases over time since some suspension is left behind, coating the wall of the capillary tube. We then compute the velocity of the front and rear menisci, $u_{\rm f}={\rm d}x_{\rm f}/{\rm d}t$ and $u_{\rm r}={\rm d}x_{\rm r}/{\rm d}t$, respectively [figure \ref{fig:setup}(c)]. Since ${\rm d}L/{\rm d}t = u_{\rm f} - u_{\rm r}$, we have:\cite{aussillous2000quick}
 \begin{equation}
 - \left.\frac{{\rm d} V}{{\rm d}t}\right|_{\rm plug}= -\pi R^{2}\,\frac{{\rm d} L}{{\rm d} t} \quad {\rm and} \quad 
 \left.\frac{{\rm d} V}{{\rm d} t}\right|_{\rm film}= \pi\, u_{\rm r}\left[R^{2}-(R-h)^{2}\right], 
 \end{equation}
where $h$ is the thickness of the film and $R$ is the radius of the capillary tube. Mass conservation implies that $ - \left.{{\rm d} V}/{{\rm d}t}\right|_{plug}=\left.{{\rm d} V}/{{\rm d} t}\right|_{film}$. Therefore, the value of $h/R$ is obtained through $u_{\rm r}$ and $u_{\rm f}$ with the relation
 \begin{equation}\label{eq:method_Aussillous}
 \frac{h}{R} = 1 -  \sqrt{1+\frac{1}{u_{\rm r}}\frac{{\rm d} L}{{\rm d} t}} = 1 - \sqrt{\frac{u_{\rm f}}{u_{\rm r}}}.
 \end{equation}
Since the velocity of the suspension plug is increasing during one experiment at constant $\Delta p$, we are able to cover a small range of capillary numbers ${\rm Ca}=\eta\,u_{\rm r}/\gamma$  during each experiment at an imposed pressure. We can then obtain the corresponding film thickness $h$, as shown for one example in figure \ref{fig:setup}(d). Compared to measurements of the coating thickness by direct visualization, the advantage of this method is that curvature effects due to the cylindrical cross-section of the capillary tube are not an issue. In addition, this method has been shown to provide very reliable and accurate results with homogeneous Newtonian and viscoplastic fluids.\cite{aussillous2000quick,laborie2017yield}


\subsection{Validation: coating by a Newtonian homogeneous fluid}

 \begin{figure}
\begin{center}
 \includegraphics[width= 0.5\textwidth]{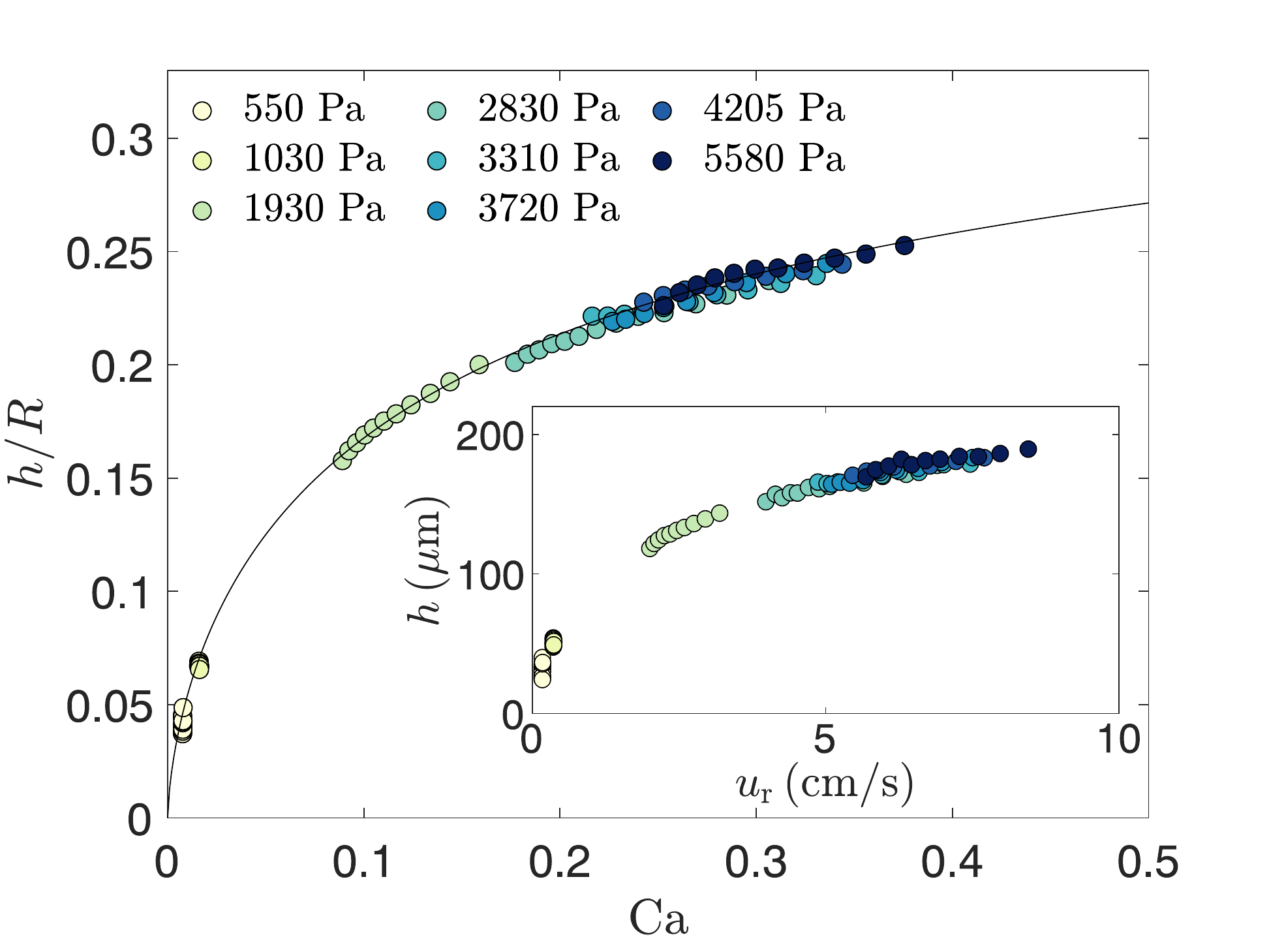}%
 \caption{Dimensionless film thickness $h/R$ as a function of the capillary number ${\rm Ca}$ for the silicone oil without particles (pure fluid, $\phi=0$). The solid line is given by Eq. (\ref{eq:Taylor}), and the circles are the experimental measurements. Each color corresponds to one experiment at a given inlet pressure $\Delta p$ (indicated in the legend). Inset: dimensional thickness of the coating film $h$ for varying velocities of the rear meniscus $u_{\rm r}$. \label{newtonian_fluids}}
 \end{center}
 \end{figure}

We first perform the experiments with the interstitial Newtonian liquid only. The film thickness left on the wall of a tube was initially predicted by Bretherton \cite{bretherton1961motion} in the limit of small capillary numbers ${\rm Ca}=\eta_{\rm f}\, U / \gamma \ll 1$, who theoretically obtained ${h}/{R} = 1.34\, {\rm Ca}^{2/3}$. The evolution of the film thickness for a broader range of parameters has been obtained through experiments and numerical simulations. In particular, Aussillous \& Qu\'er\'e \cite{aussillous2000quick} obtained experimental results consistent with the experiments carried by Taylor \cite{taylor1961deposition} and have shown that the thickness of the coating film follows the empirical law : 
 \begin{equation}\label{eq:Taylor}
 \frac{h}{R} = \frac{1.34 \,{\rm Ca}^{2/3}}{1+3.35\,{\rm Ca}^{2/3}}.
 \end{equation}
More recently, numerical simulations by Balestra \textit{et al}.\cite{balestra2018viscous} have confirmed this expression and extended it to immiscible liquids of different viscosity ratios.

The experimental data for a capillary tube of radius $R = 750\,\mu{\rm m}$ are reported in figure \ref{newtonian_fluids}. The error bars for all experiments reported in this study are typically around $\pm 5\%$ of the mean value. The experimental results match very well the prediction given by Eq. (\ref{eq:Taylor}) in the entire range of capillary numbers considered in this study, ensuring that our experimental approach and methods are correct. Besides, from the beginning to the end of an experiment, the evolution of the film thickness always follows the theoretical prediction, and there is thus no influence of the history of the system for a Newtonian fluid. We shall see later that adding particles modifies this observation.


\section{Translation of a suspension plug} \label{sec:results}

\subsection{Phenomenology}

\begin{figure}[t!]
\begin{center}
 \includegraphics[width=0.48\textwidth]{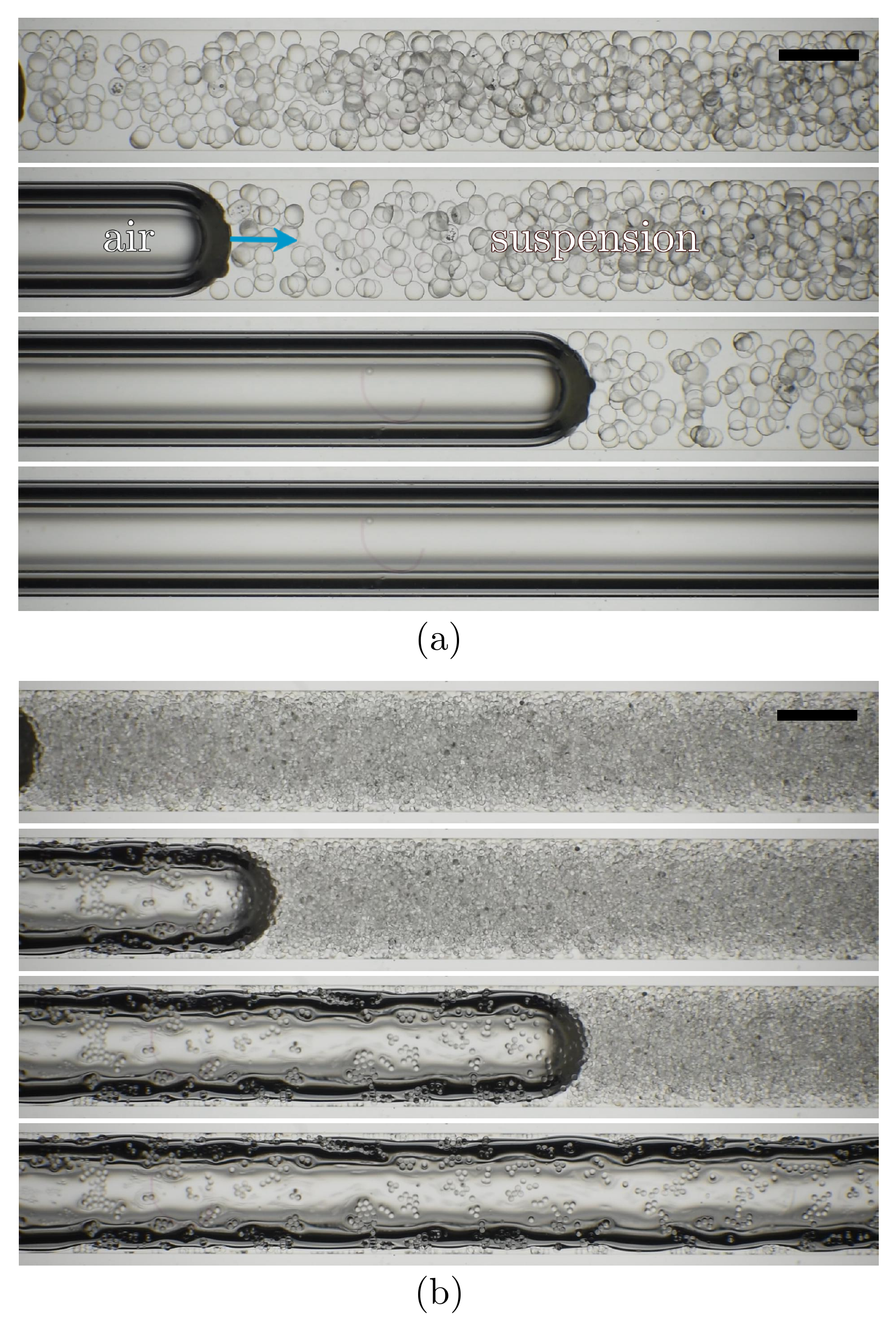}
 \caption{Examples of coating films obtained at small capillary numbers for suspensions of volume fractions $\phi=0.19$ showing (a) the "liquid-only" regime where the film formed is much thinner than the particle diameter and free of any particle, here for $d=250 \,\mu{\rm  m}$ and $u_{\rm r} \sim 0.3\,{\rm mm/s}$, and (b) the "heterogeneous film" regime for $d=80 \,\mu{\rm m}$ and $u_{\rm r} \sim 0.8\,{\rm mm/s}$ and where the coating film is thinner than the particle diameter but still allows some particles to be entrained and deposited on the outer wall of the capillary tube. The time goes from top to bottom, and the experiments are performed in a capillary tube of diameter $D=1.5\,{\rm mm}$. The arrow in (a) indicates the direction of translation of the interface, and the time goes from top to bottom. Scale bars are 1 mm. Movies are available in supplementary materials.
 \label{fig:Figure_4_OtherFilm}}
 \end{center}
 \end{figure}

\begin{figure}[t!]
\begin{center}
 \includegraphics[width=0.48\textwidth]{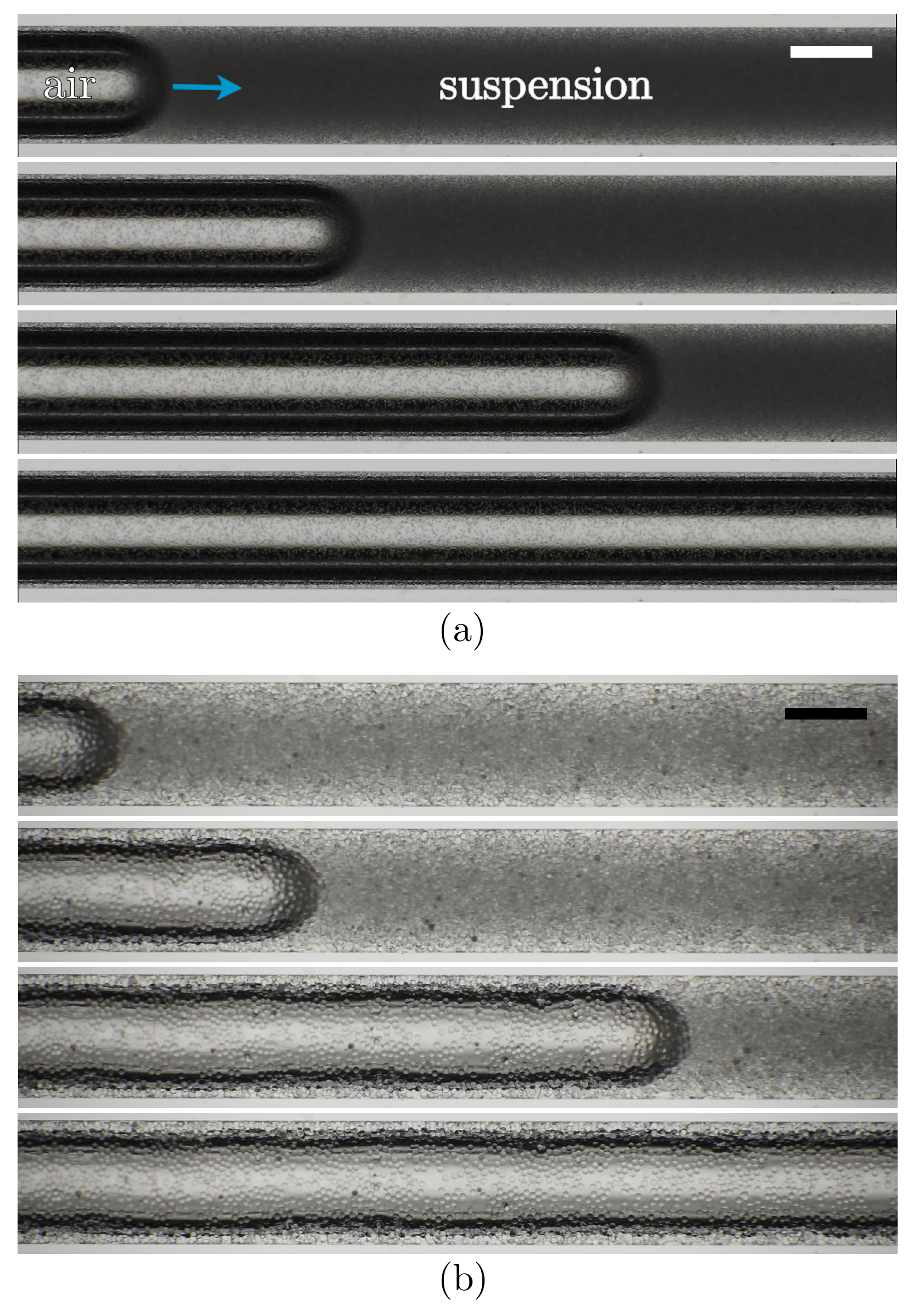}
 \caption{Examples of coating films deposited on the inner walls of a capillary tube of diameter $D=1.5\,{\rm mm}$ by a suspension of volume fraction $\phi=0.19$ of (a) $d=20 \,\mu{\rm  m}$ particles translating at $u_{\rm r} \sim 1.8\,\,{\rm mm/s}$ and (b) $d=80 \,\mu{\rm m}$ particles translating at $u_{\rm r} \sim 3.2\,{\rm mm/s}$. The arrow in (a) indicates the direction of translation of the interface, and the time goes from top to bottom. In both cases, the coating obtained is in the effective viscosity regime where a thick film made of different layers of particles is obtained. Scale bars are 1 mm. Movies are available in supplementary materials.
 \label{fig:Figure_3_ThickFilm}}
 \end{center}
 \end{figure}

 \begin{figure*}[!t]
    \centering
    \includegraphics[width=\textwidth]{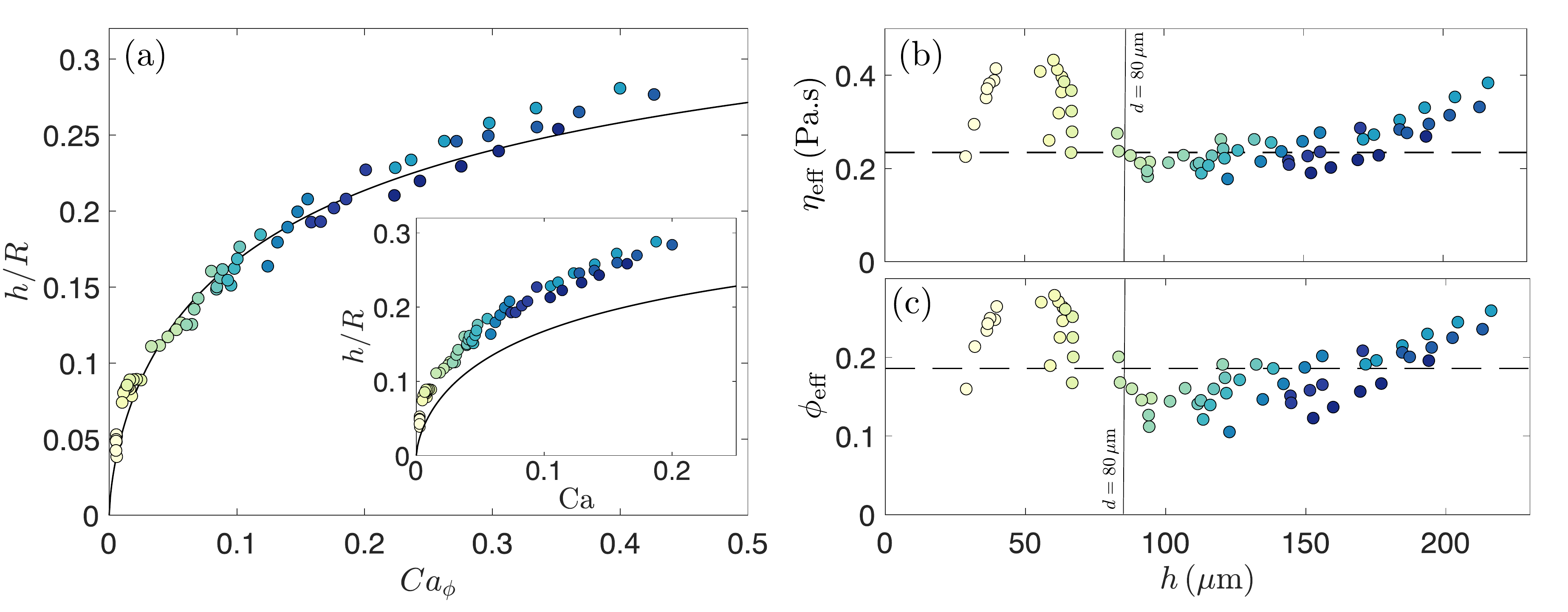}
    \caption{(a) Dimensionless deposited film thickness $h/R$ as a function of the capillary number based on the effective viscosity of the suspension ${\rm Ca}_\phi$, where the viscosity $\eta(\phi)$ is calculated with Eq. (\ref{eq:MP}) with the initial volume fraction in the syringe ($\phi=0.19$). The solid line corresponds to Eq. (\ref{eq:Taylor_phi}). Inset: $h/R$ as a function of the capillary number based on the viscosity of the interstitial fluid ${\rm Ca}$. The solid line is given by Eq. (\ref{eq:Taylor}). (b) Effective viscosity of the suspension $\eta_{\rm eff}$ leading to a coating thickness $h$ estimated using Eq. (\ref{eq:inverted}) as a function of $h$. The dotted line correspond to $\eta(\phi=0.19)=0.24\,{\rm Pa\,s}$ calculated using the Maron-Pierce correlation. (c) Estimated volume fraction of particles $\phi_{\rm eff}$ based on the Maron-Pierce correlation and obtained using Eq. (\ref{eq:MP}). The horizontal dotted line represents the initial volume fraction in the suspension, $\phi=0.19$. For all figures the experiments are performed with particles of diameter ${\rm d}=80 \,{\rm \mu m}$ dispersed at a volume fraction of $\phi=0.19$ in the silicone oil, and the plug is translated in a capillary of radius $R=750\,\mu{\rm mm}$. Each color represents an experiment performed at one given $\Delta p$.
    \label{thickness_nonnewtonian}}
\end{figure*}

We now consider the presence of neutrally buoyant and non-Brownian particles dispersed in the interstitial liquid at a volume fraction $\phi$. We report examples of the outcome of experiments performed with different particle diameters and translation velocity $u_{\rm r}$ at the same volume fraction $\phi=0.19$ in figures \ref{fig:Figure_4_OtherFilm} and \ref{fig:Figure_3_ThickFilm}. Similar to the dip coating configuration,\cite{sauret2019capillary,palma2019dip} we observe three distinct regimes. (i) At very low velocities and relatively large particle diameters, no particles are present in the coating film. The coating layer is too thin to let the particles be entrained in the film, and the particles are trapped in the meniscus which acts as a capillary filter.\cite{yu2018separation,sauret2019capillary} This first regime is referred to as the liquid only regime, corresponding to $h \ll d$, and is illustrated in figure \ref{fig:Figure_4_OtherFilm}(a). (ii) At larger velocity and relatively moderate particle diameters, the particles begin to be entrained in the coating layer as illustrated in figure \ref{fig:Figure_4_OtherFilm}(b). The particles tend to form clusters due to the capillary attraction between them to decrease the liquid/air interfacial deformation.\cite{vella2005cheerios,gans2019dip} Because of the capillary interactions between particles and the relatively low volume fraction of the suspension ($\phi=0.19$ here), the distribution of the particles in the coating layer is heterogeneous with regions of the tube inner wall covered by a liquid film only, whereas other regions are covered by a film with a monolayer of trapped particles that form clusters. The particles are randomly dispersed in the film, and the number of particles per surface area increases with the translation velocity $u_{\rm r}$. The velocity threshold between this heterogeneous regime where $h \sim d$ and the liquid-only regime depends on the particle diameter. (iii) Increasing further the velocity $u_{\rm r}$ and for small enough particle diameters, a second threshold is reached. As reported in figures \ref{fig:Figure_3_ThickFilm}(a)-(b), we observe a homogeneous coating regime consisting of a uniform coating of particles that can form multiple layers on the capillary tube wall. The suspension behaves like a homogeneous fluid with an effective viscosity modified by the presence of particles. This regime is referred to as the effective viscosity regime and corresponds to $h \geq d$. The velocity thresholds between the three different coating regimes depend on the particle diameter $d$ and the thickness of the coating film $h$. The observations in the present configuration are reminiscent of the different regimes reported in the dip coating configuration,\cite{gans2019dip} so that the strategies and models developed for the dip coating process seem to be directly applicable in the case of capillary tubes coating and deposition of thin films in porous media. However, as we shall see in the following, despite the similarities, some significant differences are present in the capillary tube configuration.

\subsection{Thickness of the coating film in the thick film regime}
 
 We experimentally measure the thickness of the coating film $h$ for a suspension of 80 ${\rm \mu m}$ particles dispersed in the silicone oil when varying the plug velocity. As before, the experiments are performed at constant pressure so that during an experiment the length of the plug decreases and the velocity $u_{\rm r}$ increases so that we obtain a series of data points at varying capillary numbers. Following the approach done for dip coating,\cite{gans2019dip} in the regime where $h \geq d$ (corresponding in the present case to $h/R \geq 0.11$) we consider that the suspension behaves as a homogeneous mixture with an effective viscosity $\eta(\phi)$. In other word, the discrete nature of the particles does not matter, and the suspension can be seen as a homogeneous fluid of larger viscosity. In this situation, we define the effective capillary number based on the viscosity of the suspension: 
\begin{equation}\label{eq:Ca_phi}
{\rm Ca}_{\phi}= \frac{\eta(\phi)\,u_{\rm r}}{\gamma}.
\end{equation} 
 The effective viscosity of the suspension can be estimated through different models such as, for instance, the Maron-Pierce correlation:\cite{guazzelli2018rheology}
\begin{equation}\label{eq:MP}
    \eta (\phi) = \eta_{\rm f} \left(1-\frac{\phi}{\phi_{\rm m}}\right)^{-2},
\end{equation}
with $\eta_{\rm f}$ the viscosity of the interstitial fluid, $\phi$ the volume fraction of the suspension, and $\phi_{\rm m}=0.59$ the maximal packing fraction for the particles and fluid used.\cite{jeong2022dip,thievenaz2022onset}

In the thick film regime, the thickness of the suspension film coating the inner wall of the capillary tube should be given by:\cite{gans2019dip,palma2019dip}
\begin{equation}\label{eq:Taylor_phi}
 \frac{h}{R} = \frac{1.34 \,{\rm Ca_\phi}^{2/3}}{1+3.35\,{\rm Ca_\phi}^{2/3}}.
 \end{equation}
 We report the dimensionless coating thickness $h/R$ as a function of the effective capillary number in figure \ref{thickness_nonnewtonian}(a) for a range of experiments performed at different $\Delta p$ and thus different $u_{\rm r}$. Here, we calculate the expected viscosity of the suspension based on the volume fraction in the syringe using Eq. (\ref{eq:MP}). Then $\eta(\phi)$ is used together with the experimental measurement of $u_{\rm r}$ in Eq. (\ref{eq:Ca_phi}) to calculate ${\rm Ca}_\phi$ and then $h/R$ using Eq. (\ref{eq:Taylor_phi}). The prediction of Eq. (\ref{eq:Taylor_phi}) captures reasonably well the evolution of the coating film despite more scattering than in the dip coating case and the capillary tube case without particles. Nevertheless, as expected, the experimental data cannot be captured by the prediction based on the viscosity of the interstitial fluid [see inset of figure \ref{thickness_nonnewtonian}(a)]. Therefore, it seems that, at first order, the results and methods obtained in the dip coating configuration in terms of ${\rm Ca}_\phi$ can be used to predict the coating of the inner wall of a capillary tube by a translating suspension plug. However, we also observe in Fig. \ref{thickness_nonnewtonian}(a) that $h/R$ tends to get larger than the prediction given by Eq. (\ref{eq:Taylor_phi}) at relatively large ${\rm Ca}_\phi$. Indeed, the data points at larger ${\rm Ca}_\phi$ are obtained near the end of an experiment when the translation velocity is the largest. The prediction becomes larger due to the non-uniformity of the coating film, as we shall discuss later.


\section{Discussion} \label{sec:discussion}

\subsection{Effective capillary number and apparent viscosity}

We have seen that Eq. (\ref{eq:Taylor}) allows predicting the evolution of $h/R$ for a homogeneous liquid.\cite{aussillous2000quick} In addition, the dip coating experiments have established that using ${\rm Ca}_\phi$ leads to a quantitative prediction of $h$.\cite{gans2019dip,palma2019dip,jeong2022dip} Since our measurements of $h$ are very accurate with the present method, we can estimate the local viscosity $\eta_{\rm eff}(\phi)$ that would lead to a coating film of measured thickness $h$ when the rear of the plug is traveling at the velocity $u_{\rm r}$.\cite{jeong2022dip} More specifically, for each experimental data point of $h$, we use Eqs. (\ref{eq:Ca_phi}) and (\ref{eq:Taylor_phi}) to obtain:
\begin{equation}\label{eq:inverted}
    \eta_{\rm eff} (\phi)=\frac{\gamma}{u_{\rm r}}\,\left(\frac{h/R}{1.34-3.35\,h/R}\right)^{3/2}.
\end{equation}
In this expression, the interfacial tension $\gamma$ is not modified by the presence of particles,\cite{couturier2011suspensions,chateau2018pinch,jeong2022dip} $R$ is a fixed parameter, and we measure $h$ and $u_{\rm r}$ so that we can calculate $\eta_{\rm eff}(\phi)$ for each data point using Eq. (\ref{eq:inverted}). In addition, once we have an estimate of $\eta_{\rm eff}(\phi)$, we can calculate the expected volume fraction $\phi_{\rm eff}$ of the film deposited on the wall of the capillary tube using the Maron-Pierce correlation [Eq. (\ref{eq:MP})]:
\begin{equation}\label{eq:invert_phi}
\phi_{\rm eff}=\phi_{\rm m}\left(1-\sqrt{\frac{\eta_{\rm f}}{\eta(\phi)}}\right),
\end{equation}
 with $\phi_{\rm m}=0.59$.

We report the estimation of $\eta_{\rm eff}$ and $\phi_{\rm eff}$ in figures \ref{thickness_nonnewtonian}(b) and \ref{thickness_nonnewtonian}(c), respectively. We have superimposed the volume fraction in the initial plug, $\phi=0.19$, and the corresponding viscosity calculated with Eq. (\ref{eq:MP}). We observe that, in average, during the entire translation of the suspension plug, the estimated viscosity and the corresponding volume fraction leading to the local coating film are of the same order of magnitude as the initial composition of the suspension. Note that in figures \ref{thickness_nonnewtonian}(b) and \ref{thickness_nonnewtonian}(c) one color of data point corresponds to a single experiment. For a given experiment, there is also some dispersion around the viscosity of the initial suspension. The same observation can also be done with $\phi_{\rm eff}$ calculated from the Maron-Pierce correlation. $\phi_{\rm eff}$ does not seem constant and seems initially smaller than $\phi=0.19$ but then increases and becomes larger at the end. These results suggest again that an effective viscosity based on the initial composition of the suspension provides a good first estimate of the resulting coating film but cannot explain the systematic deviation obtained during one experiment. We shall come back on this point later.


\subsection{Influence of the volume fraction of the suspension}
   
\begin{figure}[!t]
    \centering
    \includegraphics[width=0.5\textwidth]{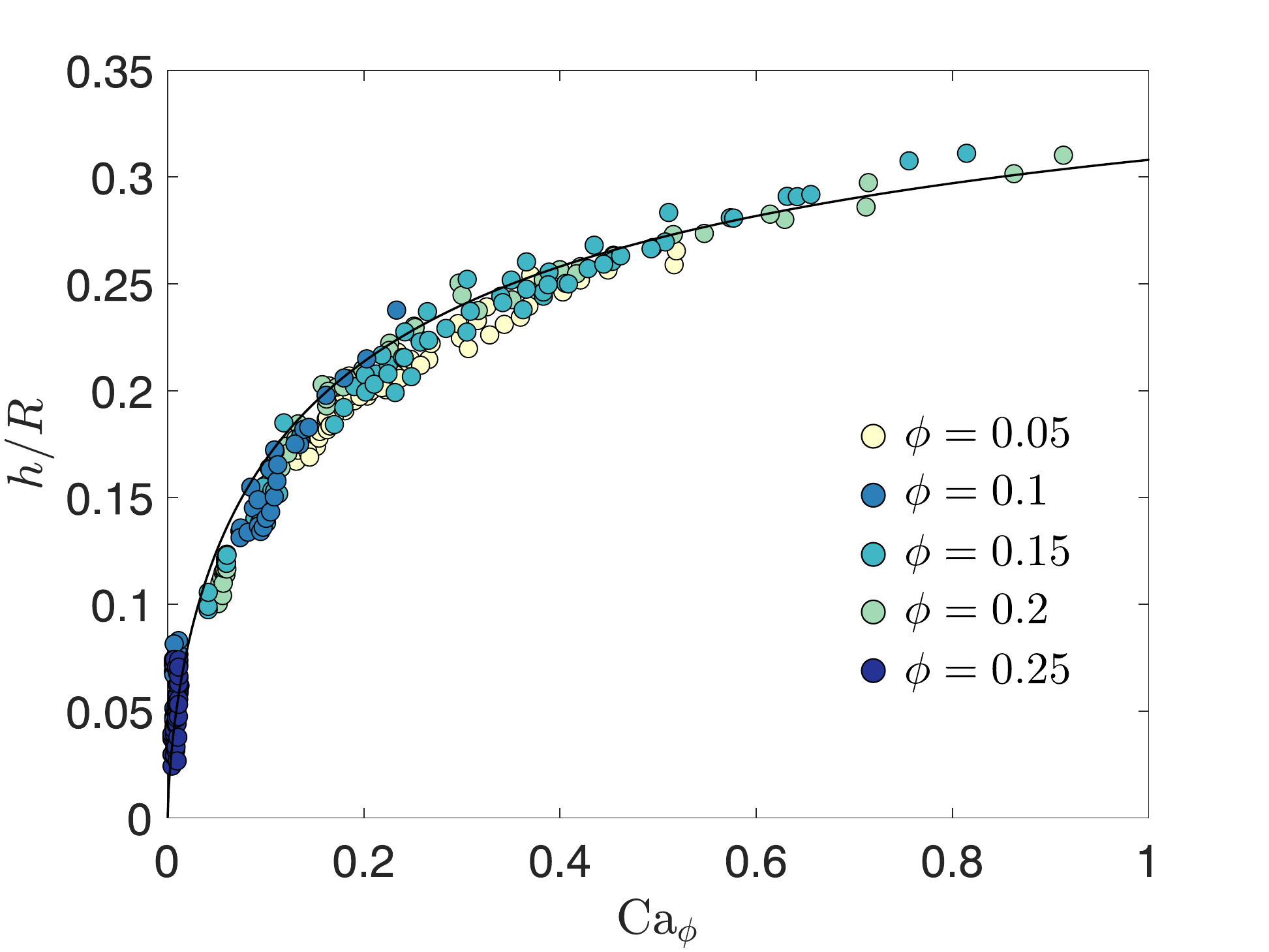}
    \caption{Rescaled thickness of the coating layer $h/R$ as a function of the effective capillary number ${\rm Ca}_{\phi}$ for varying initial volume fraction in the suspension plug, from $\phi=0.05$ to $\phi=0.25$. The experiments are performed with a suspension of 80 ${\rm \mu m}$ particles dispersed in silicone oil and the capillary tube has a radius $R=750\,\mu{\rm m}$. The solid line is given by Eq. (\ref{eq:Taylor_phi}) and the viscosity used to calculate ${\rm Ca}_\phi$ is based on the Maron-Pierce correlation with $\phi$ corresponding to the initial volume fraction of the suspension in the syringe and indicated in the legend.}
    \label{influence concentration}
\end{figure}

To provide a more exhaustive characterization, we now consider different initial volume fractions $\phi= 0.05,\, 0.1,\, 0.15,\, 0.2$, and $0.25$ for $80\,\mu{\rm m}$ particles. The results are reported in figure \ref{influence concentration}, which shows the evolution of the rescaled coating thickness $h/R$ as a function of the effective capillary number ${\rm Ca}_\phi$ and the comparison with the prediction given by Eq. (\ref{eq:Taylor_phi}). Here again, the value of $\eta(\phi)$ is calculated with the initial volume fraction in the syringe and the Maron-Pierce correlation. Apart from the very small values of $h/R <0.1$, where the coating film is thinner than the particle diameter, the experimental data matches well the prediction given by Eq. (\ref{eq:Taylor_phi}). Again, some deviations around the average values are observed, and the experimental points seem more scattered at larger volume fractions.

 \subsection{Influence of the particle diameter}

\begin{figure}[!t]
    \centering
    \includegraphics[width=0.5\textwidth]{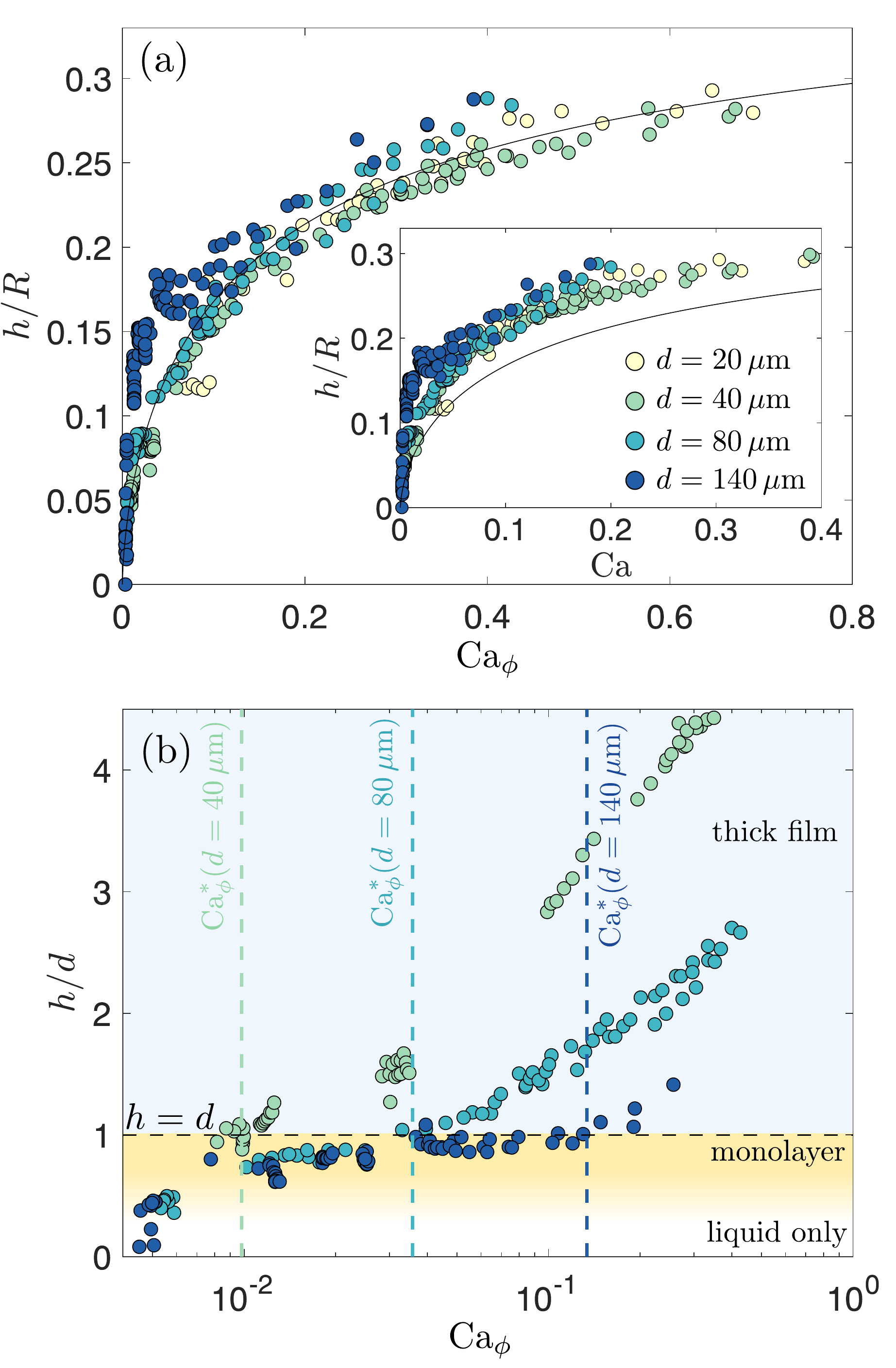}
    \caption{(a) Rescaled thickness of the coating film $h/R$ as a function of the effective capillary number ${\rm Ca}_{\phi}$ for varying particle diameters: $20 \,{\rm \mu m}$ (yellow circles), $40 \,{\rm \mu m}$ (green circles), $80 \,{\rm \mu m}$ (light blue circles) and $140 \,{\rm \mu m}$ (dark blue circles). The volume fraction of particles in the initial suspension is $\phi=0.19$, the viscosity used to calculate ${\rm Ca}_{\phi}$ is calculated using the Maron-Pierce correlation, and the inside radius of the tube is $R= 750\,\mu{\rm m}$. Inset: Evolution of $h/R$ as a function of the capillary number based on the viscosity of the interstitial fluid ${\rm Ca}$. The solid line is the coating thickness given by equation (\ref{eq:Taylor}). (b) Evolution of $h/d$ when varying the effective capillary number ${\rm Ca}_{\phi}$ and for different size of particles. The horizontal dashed line indicates $h=d$, the vertical dotted lines are the threshold effective capillary number ${\rm Ca}_{\phi}^*$ given by Eq. (\ref{eq:cond_phi}) at which the coating film starts to be in the thick film regime. The blue region denotes the regime of thick film, the yellow regime is the monolayer regime, and the white region is when only liquid is observed.}
    \label{influence_2a}
\end{figure}

\begin{figure*}[!h]
    \centering
    \includegraphics[width=0.9\textwidth]{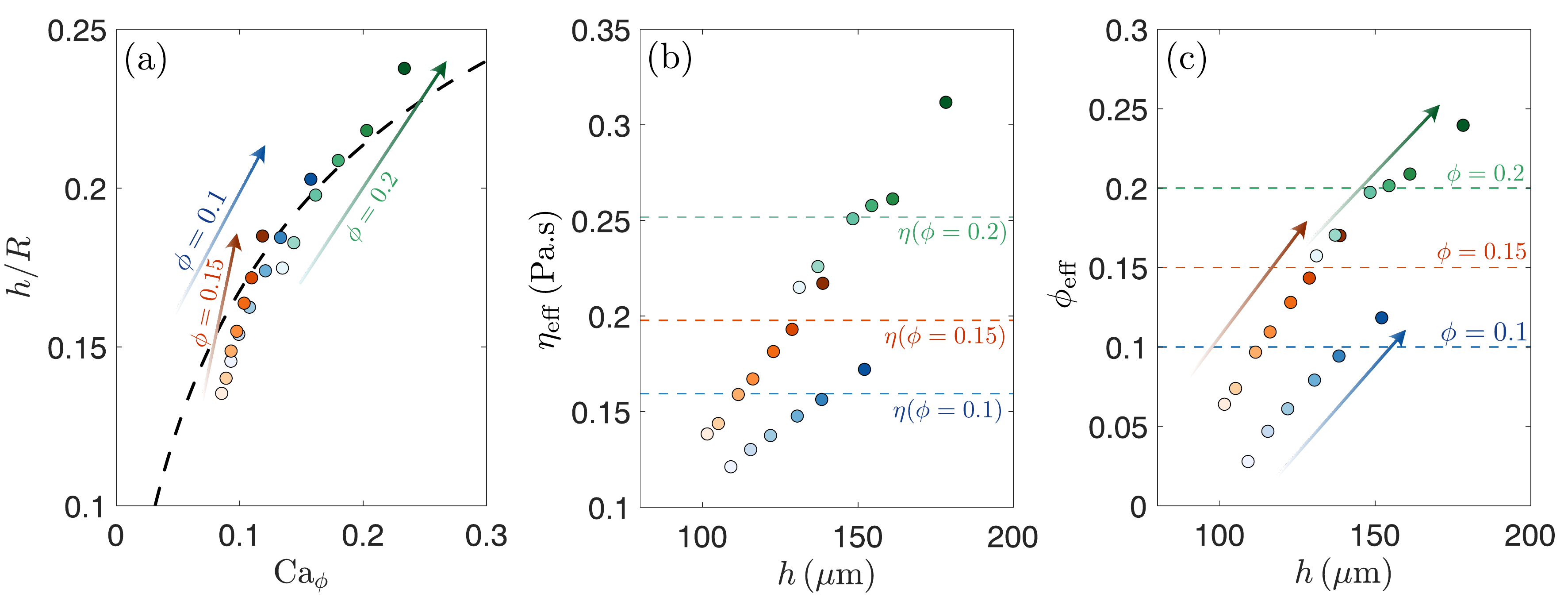}
    \caption{Examples of evolution of (a) the dimensionless coating thickness $h/R$ as a function ${\rm Ca}_\phi$, (b) the calculated corresponding effective viscosity $\eta_{\rm eff}$ and (c) the calculated corresponding solid fraction $\phi_{\rm eff}$ based on the viscosity calculated as a function of the coating thickness. The experiments are performed with a suspension of 80 ${\rm \mu m}$ particles and the capillary tube has a radius $R=750\,\mu{\rm m}$. The dotted line in (a) corresponds to Eq. (\ref{eq:Taylor_phi}) where $\eta(\phi)$ is based on the viscosity of the suspension in the syringe. Each series of colors indicates one experiment going from the light to the dark colors (the direction of time is indicated by the arrow).}
    \label{fig:shearinduced}
\end{figure*}

To investigate if the scattering of the data observed can be due to a peculiar coating regime, we consider the influence of the diameter of the particles on the coating thickness in figures \ref{influence_2a}(a)-(b). Similar to previously, we report in figure \ref{influence_2a}(a) the evolution of $h/R$ as a function of ${\rm Ca}_\phi$ (and as a function of ${\rm Ca}$ in the inset of figure \ref{influence_2a}(a)). Again, for small enough particles, the prediction given by Eq. (\ref{eq:Taylor_phi}) using $\eta(\phi)$ calculated using Eq. (\ref{eq:MP}), provides a fair estimate for the coating thickness for small and moderate sizes of particles. However, the dispersion of the points around the expected value seems to increase when increasing the particle diameter. In particular, for $d=140\,\mu{\rm m}$ particles the collapse is poor.

We rescale in figure \ref{influence_2a}(b) the coating thickness $h$ by the particle diameter $d$. Past experiments in the dip coating configuration have reported that the thick film regime is observed for effective capillary numbers such that $h \geq d$.\cite{gans2019dip,palma2019dip} A similar approach in the present situation using Eq. (\ref{eq:Taylor_phi}) together with the condition $h/d \geq 1$ leads to the threshold effective capillary number to be in the thick regime:
\begin{equation}\label{eq:cond_phi}
{{\rm Ca}_\phi}^*=\left[\frac{d}{R}\,\frac{1}{1.34-3.35\,d/R}\right]^{3/2}.
\end{equation}

We report the capillary threshold in figure \ref{influence_2a}(b) for the different particle diameters. We observe that for ${\rm Ca}_{\phi}<{{\rm Ca}_\phi}^*$ the thickness indeed seems to saturate around the particle diameter, $h \simeq d$. This observation is reminiscent of what was reported in the dip coating of a plate and corresponds to the heterogeneous regime with a monolayer of particles.\cite{gans2019dip} The larger the particle, the larger the capillary number needs to be to recover the effective viscosity regime. These results show that our experimental results quantitatively match the prediction of the threshold value ${{\rm Ca}_\phi}^*$. Beyond ${{\rm Ca}_\phi}^*$ the thickness increases following Eq. (\ref{eq:Taylor_phi}) but with some scattering of the data. This result and the threshold capillary number echo the observations made in the dip coating configuration.\cite{gans2019dip,palma2019dip}

In summary, the different coating regimes, their domains of existence, and the estimated coating thicknesses are comparable between the present case of coating of the inner wall of a capillary tube and the dip coating process.\cite{gans2019dip} At first order, this result can be expected since the two configurations share very similar fluid mechanics features. The topology of the flow in the two situations are comparable and, in particular, they both present a stagnation point followed by a coating film of constant thickness.\cite{krechetnikov2010application,jeong2022dip} Previous studies in the dip coating configuration and in the present capillary tubes configuration but with isolated spherical particles have reported that the threshold for single particle entrainment are similar.\cite{sauret2019capillary,jeong2020deposition}. In the case of non-dilute suspensions, the coating of the capillary tube can be predicted at first order using the average volume fraction, and the corresponding effective viscosity. However, as reported above, a significant difference is observed in the capillary tube configuration: during one experiment, a systematic deviation from the mean value of $h/R$ is observed. It leads to much more scattering of the data than what was observed for the dip coating. We consider in the next section this point in more detail, and in particular, we highlight the differences between the dip coating and the intermittent flow in a tube.

 \subsection{Evolution of the coating thickness and shear-induced migration}

At the beginning of an experiment, we introduce a plug of length $L$ with a controlled volume fraction $\phi$. Then, once the suspension plug translates, it deposits a suspension film on the wall of the capillary tube, making the plug shrink in volume and thus in length. We show in figure \ref{fig:shearinduced}(a) three examples of experiments at volume fractions $\phi=0.1,\,0.15$ and $0.2$ where we report the evolution of the dimensionless film thickness from the beginning of the experiment when the plug is long (light colors) to the end of the experiment (dark colors) when the plug is significantly shorter. In all cases, we observe the same evolution of $h/R$. In average, the coating thickness has a mean value centered around the prediction given by Eq. (\ref{eq:Taylor_phi}) based on the viscosity $\eta(\phi)$ calculated with the initial volume fraction. However, $h/R$ is smaller than the mean value at the beginning of an experiment and larger at the end. This effect is peculiar to the present configuration in a tube and was not observed for dip coating.\cite{gans2019dip} To shed light on this peculiar behavior, we extract the local viscosity $\eta_{\rm eff}$ when the film is deposited in figure \ref{fig:shearinduced}(b) and the corresponding volume fraction $\phi_{\rm eff}$ in figure \ref{fig:shearinduced}(c). Assuming that the suspension behaves as a homogeneous mixture, we estimate the  effective viscosity of the suspension $\eta_{\rm eff}$ leading to a coating thickness $h$ using Eq. (\ref{eq:inverted}).\cite{jeong2022dip} Then, we estimate the volume fraction of particles $\phi_{\rm eff}$ using Eq. (\ref{eq:invert_phi}). It appears that the volume fraction of particles in the coating film increases during an experiment. It is initially smaller than the volume fraction in the syringe and becomes larger at the end of the experiment when the suspension plug has shrunk significantly.

The flow of a suspension plug in a capillary tube presents a key difference with the dip coating: the shear-induced migration of the particles,\cite{leighton1987shear,lyon1998experimental,rashedi2020shear} and a "fountain-flow" where particles are more concentrated at the front meniscus.\cite{karnis1967flow,kim2017formation,luo2018particle,bouhlel2019convective,chen2021self} Indeed, the flow of a plug in a cylindrical channel is a shear flow where the particles migrate from high shear stress to low shear stress region.\cite{leighton1987shear} As a result the volume fraction of particles is not constant along a cross-section of the tube, but instead larger at the center and smaller near the outer wall. As a result, the measured local viscosity of the suspension also changed spatially. The migration of the particles is the result of combined effects of hydrodynamic forces between the particles and non-hydrodynamic interactions such as surface friction or repulsive forces between particles. Different models have been developed to predict the distribution of particles, such as the diffusion model,\cite{phillips1992constitutive}, and the suspension balance model.\cite{nott1994pressure,morris1999curvilinear} However, the present configuration exhibits an additional challenge, the plug is limited in length compared to an infinite Poiseuille flow. Therefore, a "fountain flow", with particles at the center going faster than the one near the boundary, leading to an internal recirculation is also observed.\cite{karnis1967flow} In particular, in addition to being non-uniform along a cross-section, the volume fraction is also not uniform along the length of the suspension plug.\cite{bouhlel2019convective} A theoretical description of the distribution of particles at every location in the suspension plug is beyond the scope of the present study given the high complexity of the problem. Nevertheless, we can provide a qualitative explanation for the evolution of the film thickness during an experiment.

Due to the shear-induced migration and the fountain flow, the volume fraction of particles near the outer wall at the rear of the plug is initially smaller than the average volume fraction $\phi$ prepared in the plug. Since the suspension film initially deposited on the wall comes from the streamlines near the outer wall (see \textit{e.g.}, Ref. \cite{jeong2020deposition} for a description of the flow), the initial coating film also has a smaller volume fraction, as observed in figure \ref{fig:shearinduced}(c). Therefore, the volume fraction in the remaining suspension plug increases in time. As a result, even if the volume fraction of the suspension near the wall remains smaller than near the centerline the total volume fraction in the suspension plug increases significantly during an experiment. Consequently, later in one experiment the volume fraction $\phi_{\rm eff}$ in the region of thickness $h$ near the wall increases in time, and eventually becomes larger than the initial volume fraction, as can be seen in the late stage of all examples shown in figures \ref{fig:shearinduced}(a)-(c).

One important consequence of this effect is that, even if in average it is possible to deposit a film of thickness and composition of the same order of magnitude as the initial composition of the suspension, the actual thickness and composition will both increase along the tube during the translation of the plug. Therefore, although this method can be interesting to deposit particles on the inner wall of a capillary tube,\cite{hayoun2018method} it may be challenging to control very well the resulting coating and properties. We should emphasize that such an effect would also be observed at a constant flow rate, where $u_{\rm r}$ would be constant, but where $\phi$ would follow a similar evolution.


\section{Conclusion} \label{sec:conclusion}

In summary, despite the shear-migration of particles leading to a non-homogeneous particle distribution in a cross-section of the tubing, our results suggest that for significant coating thickness, larger than the diameter of the particles ($h>d$), a good estimate of the film thickness is given by equation (\ref{eq:Taylor_phi}). In this expression, the usual classical capillary number has to be replaced by a capillary number based on the effective viscosity of the suspension [Eq. (\ref{eq:Ca_phi})]. This observation is reminiscent of the observation obtained with the dip coating configuration.\cite{gans2019dip,palma2019dip} More specifically, the same three coating regimes are recovered: (i) thin-film of pure liquid only, (ii) a heterogeneous regime where the thickness is almost constant, $h \sim d$ over a range of capillary number, and (iii) a thick-film regime consisting of multiple layers of particles. Yet, in this third regime, the presence of local variation of volume fraction leads to an evolution of the coating thickness and composition of the liquid film during the translation of the plug. Nevertheless, the mechanisms of capillary filtration and sorting proposed in the dip coating case should still mostly hold in the present configuration.\cite{sauret2019capillary,dincau2019capillary,jeong2022dip}

However, a main difference highlighted by our experiments is a systematic evolution of the coating thickness, and the local volume fraction, during the translation of the plug. This systematic evolution is associated with the complex flow in the plug: a combination of shear-induced migration and fountain flow. As a result, even if at first order the average volume fraction in the coating is the one of the plug, it is initially smaller and at later stages larger. It leads to two main consequences; First, the composition of the coating film and the thickness changes along the capillary tube. This result would still be observed for a constant translation velocity. Second, contrarily to a Newtonian homogeneous fluid, there is a strong influence of the history of the system. Indeed, the evolution of the volume fraction is going to depend on the volume of suspension already left on the wall of the tube. 

 A more refined theoretical model that captures the observed discrepancies due to the inhomogeneity of particles in the plug needs to be developed. Such a model could contribute to rationalizing the experimental results at the next order and account for the local volume fraction around the air-liquid meniscus. This prediction would also allow a uniform coating on the wall of capillary tubes by varying the velocity of the plug, for instance, in a flow-rate driven system, to account for the variation of thickness and volume fraction. Note that the coating films observed in the present configuration typically contain only a few layers of particles so that confinement could introduce additional effects. For instance, since the interface exerts a capillary force on the particles, force chains could be generated between particles leading to second-order effects. In addition, the presence of the interface and the local curvature could modify the local volume fraction of particles, as considered recently during the motion of a contact line.\cite{zhao2020spreading,pelosse2022probing}
 
 We should also emphasize that the variation in volume fraction during the translation of a suspension plug in a porous media could also favor clogging by bridging at constrictions.\cite{dressaire2017clogging,marin2018clogging,vani2022influence} The present studies illustrate the heterogeneities and complexity brought by the presence of particles in capillary systems where the size of the particle is comparable to the capillary object.

\section*{Acknowledgements}
This material is based upon work supported by the National Science Foundation under NSF Faculty Early Career Development (CAREER) Program Award CBET No. 1944844 and by the American Chemical Society Petroleum Research Fund through the ACS-PRF 60108-DNI9 grant.

\balance

\bibliography{Bretherton_Biblio} 
\bibliographystyle{ieeetr} 

\end{document}